\documentclass[9pt]{article}
\pdfoutput=1
\usepackage[utf8]{inputenc}
\usepackage{amsmath}
\usepackage{amsfonts}
\usepackage{amssymb}
\usepackage{lipsum}
\usepackage{stfloats}
\usepackage{subfig}
\usepackage[english]{babel}
\usepackage[
backend=biber,
style=alphabetic,
]{biblatex}

\usepackage[letterpaper,top=2cm,bottom=2cm,left=3cm,right=3cm,marginparwidth=1.75cm]{geometry}
\usepackage{graphicx}
\usepackage[colorlinks=true, allcolors=blue]{hyperref}
\twocolumn
\title{Automatically layout and visualize the biological pathway map with spectral graph theory}
\author{Lingran Xiao \\ lx20@rice.edu \and Yanfei Wang \\ mintywangtvt@gmail.com \and Shiying Li \\ shiyingli7-c@my.cityu.edu.hk \and Lingxi Chen \\ chanlingxi@gmail.com \and Shuaicheng Li \\ shuaicli@cityu.edu.hk}

\date{August 2021}
\begin{document}
\maketitle
\begin{abstract}

The pathway is a biological term that refers to a series of interactions between molecules in a cell that cause a certain product or a change in the cell \cite{NIH2020}. Pathway analysis is a powerful method for gene expression analysis \cite{Elisa2017}. Through pathway maps, the lists of genes that are differently expressed across the given phenotypes are translated into various biological phenomena \cite{Nguyen2019}. 

Visualizing a pathway map manually is a common practice nowadays because of the limitations of existing solutions to draw complicated graphs (i.e. directed graphs, graph with edge crossings, etc).

This project provides a solution to draw pathway maps automatically based on spectral graph theory and topological sort. Various methods are taken to enhance pathway maps' readability. Significant reductions in the number of edge crossings and the sum of adjacent nodes are achieved.

\end{abstract}

\section{Introduction}
The pathway is a biological term that refers to a series of interactions between molecules in a cell that cause a certain product or a change in the cell \cite{NIH2020}. Pathway analysis is a powerful method for gene expression analysis \cite{Elisa2017}. Through pathway maps, the lists of genes that are differently expressed across the given phenotypes are translated into various biological phenomena \cite{Nguyen2019}. 

Unlike other types of data visualization, in the node-link graphs, nodes’ positions have no intrinsic meaning and are completely determined by convenience and readability. The layout with qualified readability should have the following characteristics: uniform distribution of nodes and edges, minimized edge-crossings, minimized edge bending ratio, minimized edge lengths, and so forth \cite{Tarawneh2012}. Several node-link graphs' layout algorithms will be introduced here, however, they all fail to generate pathway maps.

Force-directed layout, which is also known as the spring layout algorithm, is among the most popular layout algorithms for node-link diagrams \cite{Tarawneh2012}. Each node is assumed to possess both attraction forces and repulsive forces.

The attraction force $f_a$ is assigned to adjacent nodes connected by a link, while the repulsive force $f_r$ is assigned to every graph node. In this way, the total energy of the spring system could be reduced by rearranging the nodes’ coordinates. The definitions of forces are as follows according to R. M. Tarawneh, et al.

$f_a(d) = k_alog(d),    f_r(d) = k_r/d^2$.

where $k_a$ and $k_r$ are constants and d is the current distance between two nodes.(“attraction forces and repulsive forces”, 2011, p153) \cite{Tarawneh2012}.

The force-directed algorithm is simple to operate and is capable to produce symmetric layouts. However, it’s regarded as an expensive algorithm, especially with a large graph’s size. The time complexity of the force-directed algorithm exceeds $O(n^3)$ with n denoting the number of graph nodes. Besides, despite force-directed algorithm functions well with undirected graphs, it could not be utilized to generate directed graphs, which means that pathway graph could not be generated through the force-directed algorithm. Moreover, force-directed layouts lack predictability. A different layout could be obtained if the algorithm is run the second time. Consequently, difficulty occurs to maintain the user’s intended force-directed graph.

Planar graphs are graphs that can be drawn in linear time without edge crossings \cite{Tarawneh2012}. The drawing of a planar graph could be realized through various algorithms, one of which is algorithm FPP. However, to apply the algorithms, several prerequisites must be satisfied, including the possibility to draw the planar graph without edges crossings. Therefore, these algorithms' application is limited. They could not generate pathway maps, which are highly possible to contain edge crossings.

These common graph layout algorithms are not capable to draw pathway maps. The force-directed algorithm could not handle directed graphs. The planar graphs algorithms could not draw graphs with edge crossings. Consequently, visualizing a pathway map manually is a common practice nowadays.

In this project, spectral graph theory will be used to locate nodes properly. Spectral graph theory studies the properties of the Laplacian matrix to calculate the positions of nodes. Through spectral graph theory, the distance between nodes sharing the same edge could be minimized and consequently, the number of crossings could be minimized and graph's readability could be facilitated. After the application of spectral graph theory, nodes of the graph will be further adjusted to make the layout clearer and more standardized. For example, the edge crossings will be eliminated, the edge lengths will be minimized, the edge-node overlapping will be avoided and so forth.

This project provides an automatic way to generate a pathway map. Comparatively, placing nodes at appropriate positions manually is time-consuming, especially with a large number of nodes and edges. 

With the input of nodes and edges, the positions of nodes could be calculated and a clarified layout could be automatically displayed. Here is an example of the intended outcome \cite{KanehisaLaboratories2020}:

\begin{figure}[h]
\centering
\includegraphics[width=0.5\textwidth]{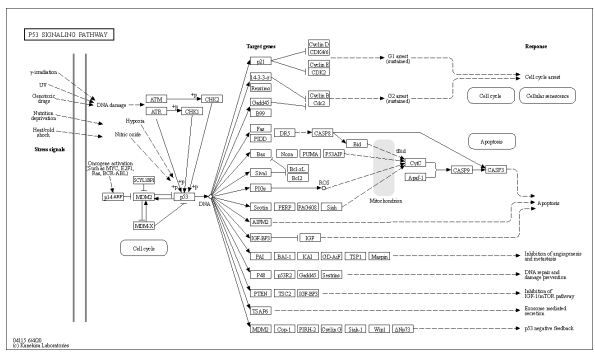}
\caption{\label{fig:1-4}Example of pathway map}
\end{figure}

\section{Methodology}

\subsection{Calculation of coordinates}

\subsubsection{data input}

\begin{figure}[h]
\centering
\includegraphics[width=0.5\textwidth]{figures/node_sample.pdf}
\caption{\label{fig:1-4}Example of input to reveal nodes' information}
\end{figure}

\begin{figure}[h]
\centering
\includegraphics[width=0.4\textwidth]{figures/edge_sample.pdf}
\caption{\label{fig:1-4}Example of input to reveal edges' information}
\end{figure}

Figure 2 and figure 3 presents the data input. Figure 2 presents the sample of nodes' description. To describe a node, five attributes are required. The first is a number used to refer to the node. The second one to the fifth one are the attributes used to visualize nodes: including shape, texts, the color of node, and the color of texts. A special design is implemented to denote node complex. Integers refer to the complex while the float number refer to single nodes belonged to complexes. If an integer is equal to the number rounded down by a float, then the integer represents this node's complex node.

Figure 3 presents the sample of edges' description. The first two attributes refer to the two adjacent nodes. The third attribute refers to the style of line. The fourth attribute refers to the style of the arrow. The fifth attribute refers to the line script. The sixth attribute refers to the additional segment of the line, for example, to represent dissociation in protein-protein interactions, a short bar perpendicular to the line is required. 

\subsubsection{Spectral graph theory}

Some basic concepts of graph theory will be explained, including the definition of adjacency matrix, degree matrix, Laplacian matrix, and so forth. The definitions 3.1 to 3.3 are cited from (Jiang ,2012) \cite{Jiang2012}.

Definition 3.1. A graph is an ordered pair G = (V, E) of sets, where

$E \subset\{\{x,y\}|x,y\in V, x\neq y\}$
(“graph”, 2012, p1) \cite{Jiang2012}.

Definition 3.2. In a graph G = (V, E), two points $x_i$, $y_j$ are adjacent or neighbors if $\{x_i,y_j\}\in E$.(“adjacent”, 2012, p2) \cite{Jiang2012}.

Definition 3.3. In the Laplacian matrix $L_G$ of the graph G, the entries $l_{i,j}$ are
\begin{equation}
  l_{i,j} =
    \begin{cases}
      -1 & \text{if {i,j} $\in E $}\\
      d(i) & \text{if i = j, and}\\
      0 & \text{otherwise}
    \end{cases}       
\end{equation},
where the degree d(v) of a vertex v is the number of vertices in G that are adjacent to v. (“Laplacian matrix”, 2012, p2) \cite{Jiang2012}.

Spectral graph theory studies some properties of eigenvalues of the Laplacian matrix. It’s proved that

$\vec{x}^{T}L_{G_{\{u,v\}}}\vec{x} = (x_u - x_v)^2$ for all $\vec{x} \in \mathbb{R}$ \cite{Jiang2012}.

Therefore, $\vec{x}^{T}L_{G}\vec{x} = \sum\limits_{{\{u,v}\} \in E} (x_u - x_v)^2$ for all $\vec{x} \in \mathbb{R}$ 

Since $\vec{x}^{T}L_{G}\vec{x} = \vec{x}^{T}(\lambda\vec{x})
= \lambda(\vec{x}^{T}\vec{x})$,

 $\lambda(\vec{x}^{T}\vec{x}) = \sum\limits_{{\{u,v}\} \in E} (x_u - x_v)^2$.
 
SVD is used to calculate the coordinates of nodes through spectral graph theory. It takes a matrix as an input (defined as A, where A is a n x p matrix). The SVD theorem is 

$A_{nxp} = U_{nxn}S_{nxp}V^T_{pxp}$,

Where $U^TU = I_{nxn}, V^TV = I_{pxp}$, the columns of U are the left singular vectors; S has singular values and is diagonal, and VT has rows that are the right singular vectors \cite{Khademhosseini2002}.

With the rectangular matrix A, SVD returns the eigenvalues of A$A^T$ and eigenvectors of A$A^T$ and $A^T$A. The eigenvectors of $A^T$A make up the columns of V, the eigenvectors of A$A^T$ make up the columns of U. Also, the singular values in S are square roots of eigenvalues from A$A^T$ or $A^T$A and are arranged in descending order \cite{Hadrienj2018}.

With the application of spectral graph theory, the coordinates of nodes could be determined and the sum of the distance between adjacent nodes could be minimized. Firstly, U, S, V are obtained by SVD with laplacian matrix as input. The dot product of the smallest two positive-eigenvalue and U will be the coordinates of the graph's nodes. The smallest distance could obtained because $ \lambda(\vec{x}^{T}\vec{x}) = \sum\limits_{{\{u,v}\} \in E} (x_u - x_v)^2$ and the smallest two eigenvalues are chosen.

\subsubsection{Topological sort}

Since pathway map is a molecular interaction/reaction network diagram, directed edges are required to reflect causal-effect relationships. For clear interpretation of the interaction series, it's more well-structured to assign nodes' depth as x-coordinate. Topological sort is utilized to obtain linear ordering (i.e. node depth) of vertices. 

A Virtual node is set as the root node and is adjacent to all real nodes. Consequently, all nodes' depths are comparable. Depth-first-Search is adopted to explore as far as possible along each branch before backtracking and to obtain node depth. 

\subsubsection{normalization}

With the application of spectral graph theory and topological sort, nodes' coordinates are set. However, it's observed that nodes with more edges are clustered with their adjacent nodes, while nodes with only one edge are placed at the margin of the graph and are far away from its adjacent edge. Consequently, nodes’ overlapping occurs and the distances between adjacent nodes are highly unequal. To solve the problem, the x and y coordinate of nodes are reassigned by the sequence of nodes’ x-coordinate and y-coordinate. In this way, the distance between adjacent nodes is more homogenized and nodes' overlapping is avoided. 

\subsection{Adjustment of nodes' coordinates}

\subsubsection{Align adjacent nodes}

To align nodes horizontally and vertically, an algorithm is designed to provide every node with alternatives that have the same x coordinate or y coordinate with its adjacent node to choose from. Through this approach, nodes are aligned with their adjacent nodes to the largest extent, either horizontally or vertically. In the meantime, graph width and height could be reduced because of the reduced number of x/y coordinates. Nodes’ original position is also included as an alternative to deal with the scenario that all other alternatives are occupied.

Approaches to assign nodes with alternatives vary, depending on the number of nodes' adjacent nodes. For nodes with only one edge, there are 4 possible alternatives including its original position. The possible alternatives are shown below in Figure 4, with dashed rectangles representing alternatives despite the original one. With (x1, y1) denoting the position of the concerned node and (x2, y2) denoting the position of the adjacent node, the coordinates of the other three alternatives are (x2, y2+|y1-y2|), (x2, y2-|y1-y2|), (x1, y2) respectively. 

\begin{figure}[h]
\centering
\includegraphics[width=0.3\textwidth]{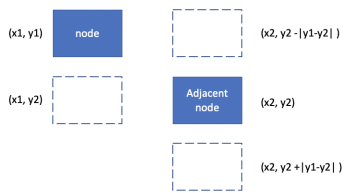}
\caption{\label{fig:fig1}possible alternatives of node with one adjacent node.}
\end{figure}

For nodes with two adjacent nodes, 3 alternatives including the node’s original position. The coordinates of alternatives are combinations of x-coordinate of concerned node and y-coordinate adjacent node1 or adjacent node2. With (x1, y1) denoting the position of the concerned node, (x2, y2) denoting the position of adjacent node1 and (x3, y3) denoting the position of adjacent node2, the coordinates of the alternatives are (x1, y1), (x1, y2), (x1, y3) respectively. In this way, the x-coordinates of node 1, node2 and node 3 is kept, aiming to reflect the causal-effect relationship of biological reactions.

\begin{figure}[h]
\centering
\includegraphics[width=0.25\textwidth]{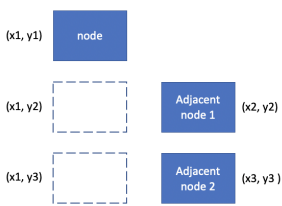}
\caption{\label{fig:fig2}possible alternatives of node with two adjacent node.}
\end{figure}

With alternatives obtained, an alternative selection algorithm will be implemented. The intention of selection is to prevent nodes' overlapping and to homogenize the layout. For nodes with one or two adjacent nodes, the selection is based on "neighbor nodes". "Neighbor nodes" are chosen based on the positions of the other adjacent nodes of the adjacent node of the concerned node, which could be denoted as "secondary adjacent nodes". The nodes that are not secondary adjacent nodes but are near to secondary nodes are also taken into consideration. The “neighbor nodes” are defined as nodes whose x/y coordinate is between the minimum x-coordinate/y-coordinate and the maximum x-coordinate/y-coordinate of secondary nodes. A margin is also added to expand the considered region and to include more neighboring nodes. The alternative that maximizes the distance from "neighbor nodes" will be chosen.

In terms of nodes with more than two adjacent nodes, the strategy to obtain alternatives is different. Thanks to spectral graph theory, the nodes with more than two adjacent nodes are often placed at the center of their adjacent nodes. Therefore, choosing the alternative with minimal change of y coordinate is adequate to align the nodes and keep the graph homogenized.

\subsubsection{Drag adjacent nodes}

It’s observed that some nodes are unnecessarily far from each other. Dragging nodes horizontally/vertically towards their adjacent nodes helps to minimize the length of edges. Nodes will be dragged in four directions, i.e., towards the largest x-coordinate, towards the smallest x-coordinate, towards the largest y-coordinate, towards the smallest y-coordinate. When conducting the dragging process, special attention should be attached to nodes that are adjacent and aligned. These nodes should continue to be aligned and share the same x/y coordinates. 

Taking dragging horizontally towards the largest x-coordinate for an example, firstly, columns, which refer to sets of nodes sharing certain x-coordinate, are determined. The number of columns is equal to the number of the x-coordinates. Then, adjacent nodes in the same column will be grouped, so that they could be treated as a single component and move simultaneously.

After the grouping process, the next step is to determine each move's destination. One possible implementation is to increase x coordinate by one recursively until an “obstacle” is found. The term “obstacle” represents a position occupied by existing nodes. However, the outcome of this implementation is not satisfactory. It could not handle the scenarios where no obstacles will be met. Under this situation, the group will move to the right end. A limit of moving should be considered.

The practical way is to determine the maximum increase of x-coordinate in advance of the movement of every group. It could be assumed that drag forces exist between adjacent nodes. Every link between adjacent nodes could be regarded to have two statuses, taut or slack. If the difference of x-coordinates between the node and its adjacent node is one, then the link could be regarded as slack in the x-direction and no further increase of x-coordinate is necessary. However, if the difference is larger than one, this link is tight in the x-direction, and nodes on the left will be under the dragging force to move to the right.

The intended x-coordinate is calculated by summing up the difference of x-coordinates between the node and every adjacent node of this node if the adjacent node’s x-coordinate is larger than the concerned node’s x-coordinate plus one. After maximum moving steps are obtained, nodes will move towards the right direction by one recursively until the moving step is reduced to 1 or until an “obstacle” is encountered. As for the grouped nodes, the tension of the group is the averaged tension of all effective links of all grouped nodes.

\subsubsection{Reduce edge crossings}

The number of edge crossings is among the most important aspects of the evaluation of a graph's layout. To reduce edge crossings, the first step is to identify edge crossings. Firstly, the two line segments are considered as two rectangles with the line segments as diagonals. If the two rectangles do not overlap, then these two line segments must not intersect. Then a straddle experiment will be conducted. If two line segments intersect, they must straddle each other, which means the two endpoints of one line segment must be on the two sides of the other line segment. Figure 6 displays two intersected line segments, where A, B are two endpoints of line 1 and C, D are two endpoints of line2. Since vectors $\overrightarrow{AD}$ and $\overrightarrow{BD}$ are on different sides of vector $\overrightarrow{CD}$, their cross products are of different signs, i.e. $\overrightarrow{AD}\cdot \overrightarrow{CD}\times\overrightarrow{BD}\cdot \overrightarrow{CD}<=0$. It should also be proved that point C and point D are on different sides of line segment AB.

\begin{figure}[h]
\centering
\includegraphics[width=0.3\textwidth]{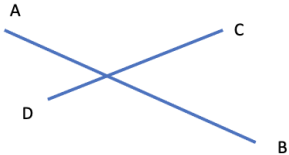}
\caption{\label{fig:fig3} an example of edge crossings}
\end{figure}

With all edge crossings identified, the next step is to sort edges by their number of edge crossings from the most to the least. For each edge, various adjustments are possible. As shown in figure 7, every adjacent node of this edge has three possible alternatives. With (x1, y1) denoting the position of node 1, (x2, y2) denoting the position of node 2, the coordinates of the alternatives are (x1, y2), (x1, y2 - 1), (x1, y2 + 1), (x2, y1), (x2, y1 - 1), (x2, y1 + 1) respectively. The units of x1, y1 are the x-sequence/ y-sequence instead of pixel. In this way, the topological sequence between node 1 and node 2 are kept and the distance between these two adjacent nodes are minimized. Consequently, some edge crossings will be eliminated. The reason is that, the y-coordinates of edge crossings' crossing points must range between these two nodes' y coordinates. If the difference between the two nodes' y coordinates minimizes, less chance exists for edge crossings. With all possibilities tested, the optimal adjustment is chosen.

\begin{figure}[h]
\centering
\includegraphics[width=0.3\textwidth]{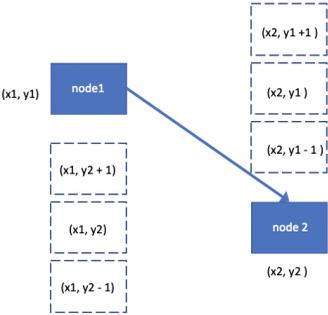}
\caption{\label{fig:fig3} possible adjustments to reduce edge crossings}
\end{figure}

In addition, since the node will move from the original position to the latter one, some nodes may be influenced by this move. For example, this node's latter position may overlap with another node. Therefore, all nodes that have same x-coordinate and whose y-coordinate ranges between the initial one and the latter one will be moved together by 1. The moving direction is decided by the moving direction of the concerned node. 

To optimize the edge reduction process, recursion is utilized to reduce edge crossings until no more edges crossings could be reduced. Since some nodes will be moved due to the edge crossing adjustment, some nodes that are aligned horizontally will not be aligned anymore. The align and drag process need to be conducted again.

\subsubsection{Add nodes' group}

Some nodes are of the same biological definition. For example, some nodes represent final products while some represent the ingredients. Since the reaction sequence is reflected through nodes' x-coordinates, it's necessary to assign the nodes of same biological definition with the same X coordinates. The nodes of same biological definition is denoted as one group. To decide groups' x coordinate, three decision techniques are designed, which are namely "foremost", "last", "voting". The group with the "foremost" label will have the smallest x-coordinate, the group with the "last" label will have the largest x-coordinate and the group with the "voting" label will choose the most common x-coordinate of the group. The x-coordinate of a group is exclusive, which means that no other nodes will be with this x-coordinate.

When adjusting the group's x-coordinate, other nodes should also be adjusted. For the group with the "foremost" label, the group's node depth is set as 0 and the depths of all other nodes are increased by 1. For the group with the "last" label, the group's depth is assigned as the maximum depth plus 1. For the group with the 'voting' label, the depth is decided by voting. As for the nodes previously with this x-coordinate, they will be moved to the left or the right by one, voted by their adjacent nodes. 

The group attribute should always be considered when adjusting node coordinates using the algorithm described earlier. When nodes are dragged horizontally, the nodes sharing the same label should be dragged together. Moreover, the labels' exclusivity is kept, no other nodes will be assigned the x-coordinate chosen by labels.

\subsection{Visualization}

\subsubsection{Different symbols of edges/nodes}
Various node shapes will be used to represent gene products and chemical compounds, etc. Meanwhile, different arrow shapes will be adopted to represent enzyme-enzyme relations, gene expression relations, protein-protein interactions, etc. 

\subsubsection{Get nodes' sizes and coordinates}
When locating a node, nodes' widths and heights should be taken into consideration. The coordinates of every node are calculated from the sizes of nodes on its left and on its top. 

Moreover, the text lengths also influence nodes' sizes. The nodes' lengths will be the default length or the length of the longest word. The nodes' widths are decided by the number of text lines. With node length obtained, words will be inserted into one text line as much as possible. A new text line will be added if the word length exceeds the decided node length.

\subsubsection{Handle edge-node overlapping}
To obtain a clear illustration and avoid misunderstanding, edges should not overlap with nodes. Crossings between edges and nodes could be divided into two categories and be eliminated differently. The first category is for edge-node crossings where the two vertices on the edge and the overlapped vertex are with the same x/y coordinate. The second category is for the other scenarios. 

For edges of nodes with the same x/y coordinate, if they overlap with a node, they could bypass them. The arrow lines will be divided into three segments. For edges with nodes sharing the same y coordinate, the arrow lines will bypass the obstacle node from above. For edges with nodes sharing the same edge and same x coordinate, the arrow lines will bypass the obstacle node from the right. 

The example of bypassing obstacle nodes is shown in Figure 8. Starting point’s position is denoted as (x1, y1), the ending point’s position is denoted as (x2, y2) and the margin from rectangles is denoted as “margin”. To represent edges of nodes sharing the same y coordinate, these three lines’ starting position and ending position are respectively (x1, y1), (x1+margin, y1); (x1+margin, y1), (x1+margin, y2); (x1+margin, y2), (x2, y2). To represent edges of nodes sharing same x coordinate, these three lines’ starting position and ending position are respectively (x1, y1), (x1, y1+margin); (x1, y1+margin), (x2, y1+margin); (x2, y1+margin), (x2, y2). The nodes' widths and lengths are neglected here. 

\begin{figure}[h]
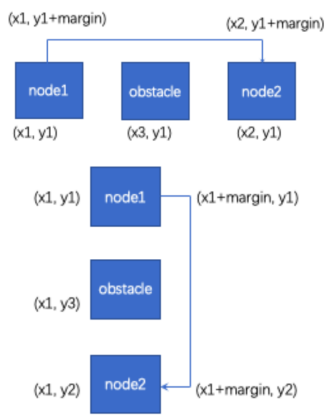

\centering
\includegraphics[width=0.3\textwidth]{figures/fig3a.pdf}
\includegraphics[width=0.25\textwidth]{figures/fig3b.pdf}
\caption{\label{fig:fig3}Methods to bypass "obstacle" node.}
\end{figure}

Another algorithm should be designed to handle the other edge-node overlapping scenarios, which are more random and unpredictable. This is realized by following steps.

Firstly, edge-node crossings are detected using edge-edge crossings with nodes represented by their four sides. Secondly, virtual nodes are initialized around the two vertices on the edge that has edge-node crossings. The node gap is set as 5 pixels. The virtual nodes adjacent in the horizontal direction and the vertical direction are connected to each other.

The third step is to detect whether the virtual node is inside a real node, if true, the edges of these virtual nodes will be eliminated. In this way, the overlapping between the concerned edge and any real node is avoided. To detect whether the virtual node's coordinate is inside the rectangle(circle could be treated as a rectangle) means to determine whether the point is on the correct side of all faces, which can be solved using cross product. With a real node represented as a rectangle with lines AB, BC, CD, DA, and a virtual node represented as a point E, both E's relationship with AB, CD and relationship with AD,BC will be detected. If E is between AB and CD, ($\overrightarrow{AB} \times \overrightarrow{AE} ) \cdot (\overrightarrow{CD} \times \overrightarrow{CE})$ should be positive. Similarly, if E is between AD and BC, ($\overrightarrow{DA} \times \overrightarrow{DE} ) \cdot (\overrightarrow{BC} \times \overrightarrow{BE})$ should be positive.

Fourthly, to minimize the number of edge bending, the possible edges connected between any face of a node and any face of the other adjacent node are to be considered. One virtual node is used to represent a node's side and four virtual nodes are used to represent a node. All possible edge paths are compared to find the optimal path. 

Fifthly, breadth-first-search is applied to find the shortest path between two concerned virtual nodes. To obtain the path, every node's predecessors are recorded. From the end node, connect it with the predecessor until the starting node is reached.  

Lastly, the optimal path among 16 shortest paths is defined as the path with fewer bending. The edge bending is represented as changes of "motion direction". The "motion direction" is decided by the relative position of the current node and the previous node. If they share the same column, the motion is decided as 'vertical'. Similarly, if they share the same row, the motion is decided as 'horizontal'. The optimal path is the path with fewer changes of motion directions. Besides, since the path is of two virtual nodes, an additional bending may be caused by the connection between the real node and the representative virtual node.

\subsubsection{bypassing lines' separation}

When handling the first scenario of edge-node crossing, some bypass lines may overlap because the margins assigned to bypass lines are the same. This kind of overlapping can lead to confusion and the level separating algorithm is necessary to assign the bypassing lines with different margin values. To accomplish this target, firstly, the bypassing edge lines are classified by their x-coordinate/y-coordinate since only bypassing edge lines with the same x-coordinate/ y-coordinate will have a chance of overlapping. The procedures of separating bypassing lines with the same x-coordinate will be explained here. The procedures of separating levels with the same y-coordinate are similar.

Inside the classification of x-coordinate, if only one bypassing edge line is included, there is no chance for this edge line to overlap with another, and the margin of this edge line is set as the default value. Else if more than one bypassing edge line is included in the classification of an x-coordinate, then a level separating algorithm should be applied.

For every classification of x-coordinate with more than one bypassing edge line, a variable 'level' should be defined to decide bypassing lines' margin. The first edge will be included in the first level. When classifying the second edge's level, an overlapping test with the first edge needs to be conducted. If its minimum y-coordinate is larger than the maximum y-coordinate of the first edge or its maximum y-coordinate is smaller than the minimum y-coordinate of the first edge, then this second edge could also be added to the first level. Otherwise, the number of levels should increase by one and the second edge should be added to the second level. This method works well for the classification of x-coordinate with two bypassing edge lines. To expand the classification method to more than two bypassing edge lines, a more detailed algorithm should be designed.

If the tested line needs to perform an overlapping test with a level including multiple bypassing lines, all included lines should be taken into consideration to determine whether the tested line could be added into the vacancies. This purpose is realized by sorting all coordinates numbers included in the level from minimum to maximum.

To add the tested edge line into a level, one of the following three conditions is needed to be satisfied. The first is that the maximum coordinate of the concerned edge should be smaller than the minimum coordinate included in this level. The second is that the minimum coordinate of the concerned edge should be larger than the maximum coordinate included in this level. The third condition is to compare the coordinates of the tested line and included lines and to find out whether the tested line could fit into the vacancy between two already included lines.

If the newly tested line couldn't satisfy these three conditions of the first level, then it will be tested for the second level and so forth until it’s tested for the highest level. If this newly tested edge could not be added to any existing level, then the number of levels will be increased by one and the tested edge will be added to the highest level. 

\subsubsection{Separating edges on the side}

Under some circumstances,edges are in the same position on nodes(e.g. the midpoint of upper face), it’s hard to distinguish the edges apart and confusion occurs. To separate the lines, the lines' position on the nodes should be different and be determined before visualization. This is realized by calculating the positions of edges on nodes in advance according to the predefined drawing faces. E.g. for nodes sharing the same x-coordinate, if no obstacles exist, the upper node’s bottom side will have one line connected, while the lower node’s above side will have one line connected.

\section{Case Study}
Figures 7, 8, and 9 presents four sample visualizations of the pathway map. The first one is with nodes randomly generated. The second one is the pathway map obtained after calculation of coordinates. The third one is the pathway map obtained after coordinates' adjustment. The fourth one is the one with edge adjustment and with edge formatting.

\begin{figure*}[h]
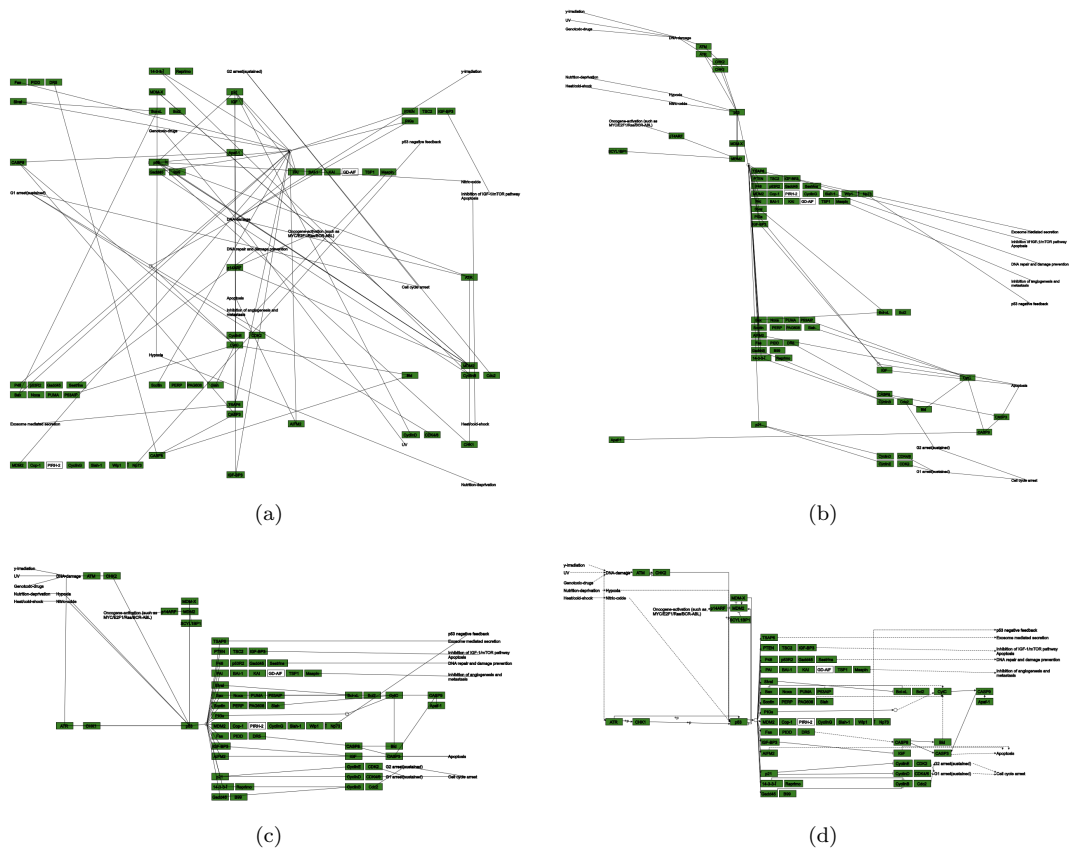

  \centering
  \subfloat[][]{\includegraphics[width=.45\textwidth]{figures/1-1.pdf}}\quad
  \subfloat[][]{\includegraphics[width=.45\textwidth]{figures/1-2.pdf}}\\
  \subfloat[][]{\includegraphics[width=.45\textwidth]{figures/1-3.pdf}}\quad
  \subfloat[][]{\includegraphics[width=.45\textwidth]{figures/1-4.pdf}}
  \caption{\label{fig:1-4}First example of pathway map.}
\end{figure*}

\begin{figure*}[h]
  \centering
  \subfloat[][]{\includegraphics[width=.45\textwidth]{figures/2-1.pdf}}\quad
  \subfloat[][]{\includegraphics[width=.45\textwidth]{figures/2-2.pdf}}\\
  \subfloat[][]{\includegraphics[width=.45\textwidth]{figures/2-3.pdf}}\quad
  \subfloat[][]{\includegraphics[width=.45\textwidth]{figures/2-4.pdf}}
  \caption{\label{fig:1-4}Second example of pathway map.}
  \label{fig:sub1}
\end{figure*}

\begin{figure*}[t]
  \centering
  \subfloat[][]{\includegraphics[width=.45\textwidth]{figures/3-1.pdf}}\quad
  \subfloat[][]{\includegraphics[width=.45\textwidth]{figures/3-2.pdf}}\\
  \subfloat[][]{\includegraphics[width=.45\textwidth]{figures/3-3.pdf}}\quad
  \subfloat[][]{\includegraphics[width=.45\textwidth]{figures/3-4.pdf}}
\caption{\label{fig:1-4}Third example of pathway map.}
  \label{fig:sub1}
\end{figure*}

\subsection{Quality evaluation}
When the figure is randomly generated, it's hard to interpret this pathway map. Firstly, the sequences of nodes are at random, making it hard to understand the series of interactions among molecules. Secondly, a huge amount of edge crossings exists, making it hard to follow the interaction logic. 

In terms of the graph obtained after coordinates calculation process, with topological sort applied, the x-coordinate of nodes is determined by their depth. Consequently, the series of interactions are reflected. With spectral graph theory applied, the distance between adjacent nodes is minimized and the layouts of pathway maps are more clear. However, it could be observed that many unnecessary vacancies exist. Besides, many adjacent nodes are not aligned horizontally or vertically.

Afterwards, multiple algorithms are applied to align nodes, reduce lengths of edges and eliminate edge crossings. The obtained pathway layout is clear and meaningful. However, more styles should be added to the edges to complete the pathway map and convey more biological meanings. For example, the solid line with a solid arrow is used to represent molecular interaction, the dashed line with a solid arrow is used to represent an indirect link or unknown reaction. Additionally, it's observed that some edge-node overlapping exist, further adjustments are necessary.

Finally, with edge-node overlapping, edge-edge overlapping handled, the final pathway maps are obtained, which are clear and meaningful.

\subsection{Quantity evaluation}
The aims of the pathway layout include: uniform distribution of nodes and edges, minimized edge-crossings, minimized edge bending ratio, minimized edge lengths, and so forth. Therefore, Figure 12(a) displays the change of edge crossings with algorithms applied. Figure 12(b) displays the sum of lengths of adjacent nodes with algorithms applied. Stage1 refers to the figure with randomly generated coordinates. Stage2 refers to the figure with spectral graph theory and topological sort applied. Stage3 refers to the figure with nodes' coordinates adjusted. It's clearly shown that the application of algorithms could reduce the number of edge crossings and the sum of the distance between adjacent nodes significantly.

\begin{figure*}[t]
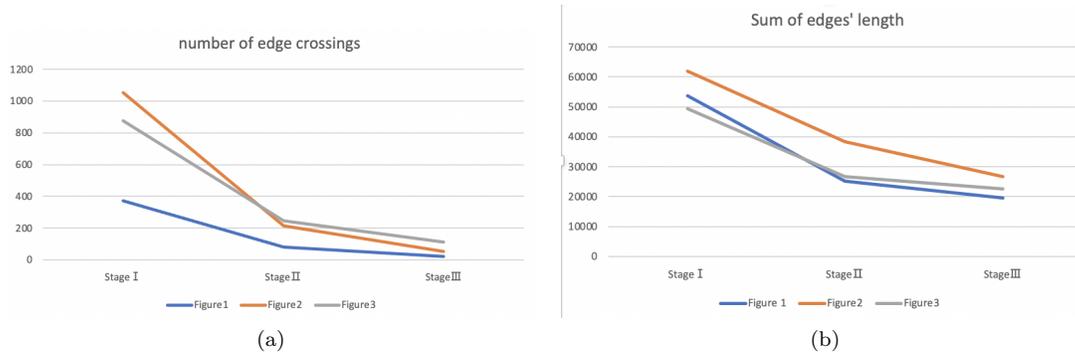

\centering
  \subfloat[][]{\includegraphics[width=0.45\textwidth]{figures/edge_crossing.pdf}}\quad
  \subfloat[][]{\includegraphics[width=0.45\textwidth]{figures/sum_edges_len.pdf}}\\
\caption{\label{fig:1-4}(a) Number of edge crossings, (b) sum of edges' lengths}
\end{figure*}

\section{Conclusion}
In a conclusion, with data input of nodes and edges, this project could visualize a graph automatically based on spectral graph theory and topological sort. Visualizing a pathway map manually is a common practice nowadays because of the limitations of existing solutions to draw complicate graphs. This project could serve as a solution to draw a pathway map automatically. After obtaining the coordinates of nodes through spectral graph theory, a layout standardizing algorithm is conducted to align nodes horizontally and vertically, homogenizing the layout, making adjacent nodes closer, and minimizing edge crossings. Methods are also taken to avoid overlapping of lines and overlapping of lines and nodes. 

In current practice, a pathway graph could be automatically displayed. However, some avoidable edge crossings may be displayed. In the future, interactions may be added for users to adjust coordinates and edges manually.

\medskip
\printbibliography

@Misc{Hadrienj2018,
  author    = {Hadrienj},
  month     = mar,
  title     = {Singular Value Decomposition},
  year      = {2018},
  url       = {https://hadrienj.github.io/posts/Deep-Learning-Book-Series-2.8-Singular-Value-Decomposition/},
}

@Misc{Jiang2012,
  author = {Jiaqi Jiang},
  month  = oct,
  title  = {An Introduction To Spectral Graph Theory},
  year   = {2012},
  url    = {http://math.uchicago.edu/~may/REU2012/REUPapers/JiangJ.pdf},
}

@Article{Tarawneh2012,
  author  = {Tarawneh, R.M. and Keller, P. and Ebert, A.},
  journal = {OpenAccess Series in Informatics},
  title   = {general introduction to graph visualization techniques},
  year    = {2012},
  month   = jan,
  pages   = {151-164},
  volume  = {27},
  doi     = {10.4230/OASIcs.VLUDS.2011.151},
}

@Misc{NIH2020,
  author = {NIH},
  month  = aug,
  title  = {Biological Pathways Fact Sheet},
  year   = {2020},
  url    = {https://www.genome.gov/about-genomics/fact-sheets/Biological-Pathways-Fact-Sheet},
}

@Misc{KanehisaLaboratories2020,
  author = {Kanehisa-Laboratories},
  month  = jun,
  title  = {PATHWAY: map04115},
  year   = {2020},
  url    = {https://www.genome.jp/entry/map04115},
}

@Misc{Khademhosseini2002,
  author = {Ali Khademhosseini},
  month  = sep,
  title  = {Singular Value Decomposition (SVD) tutorial},
  year   = {2002},
  url    = {https://web.mit.edu/be.400/www/SVD/Singular_Value_Decomposition.htm},
}

@Article{Elisa2017,
  author  = {Cirillo Elisa, Parnell Laurence D., Evelo Chris T.},
  journal = {Frontiers in Genetics},
  title   = {A Review of Pathway-Based Analysis Tools That Visualize Genetic Variants},
  year    = {2017},
  issn    = {1664-8021},
  pages   = {174},
  volume  = {8},
  doi     = {10.3389/fgene.2017.00174},
  url     = {https://www.frontiersin.org/article/10.3389/fgene.2017.00174},
}

@Article{Nguyen2019,
  author  = {Nguyen TM., Shafi A., Nguyen T. },
  journal = {Genome Bio},
  title   = {Identifying significantly impacted pathways: a comprehensive review and assessment},
  year    = {2019},
  volume  = {20},
}

\end{document}